\documentclass[twocolumn,amssymb,amsmath,aps,prl,superscriptaddress,nobibnotes,showpacs]{revtex4}
\usepackage{graphicx}
\usepackage{docs}

\begin{document}

\title{Experimental Entanglement Concentration and Universal Bell-state Synthesizer}

\author{Yoon-Ho Kim} \email{kimy@ornl.gov}
\affiliation{Center for Engineering Science Advanced Research, Computer Science
and Mathematics Division, Oak Ridge National Laboratory, Oak Ridge, Tennessee
37831}

\author{Sergei P. Kulik}
\altaffiliation[Permanent address: ]{Department of Physics, Moscow
State University, Moscow, Russia} \affiliation{Department of
Physics, University of Maryland, Baltimore County, Baltimore,
Maryland 21250}

\author{Maria V. Chekhova}
\altaffiliation[Permanent address: ]{Department of Physics, Moscow
State University, Moscow, Russia} \affiliation{Department of
Physics, University of Maryland, Baltimore County, Baltimore,
Maryland 21250}

\author{Warren P. Grice}
\affiliation{Center for Engineering Science Advanced Research, Computer Science
and Mathematics Division, Oak Ridge National Laboratory, Oak Ridge, Tennessee
37831}

\author{Yanhua Shih}
\affiliation{Department of Physics, University of Maryland,
Baltimore County, Baltimore, Maryland 21250}

\date[]{Submitted: April 16, 2002; Revised: \today}

\begin{abstract}
We report a novel Bell-state synthesizer in which an interferometric
entanglement concentration scheme is used. An initially mixed polarization
state from type-II spontaneous parametric down-conversion becomes entangled
after the interferometric entanglement concentrator. This Bell-state
synthesizer is universal in the sense that the output polarization state is not
affected by spectral filtering, crystal thickness, and, most importantly, the
choice of pump source. It is also robust against environmental disturbance and
a more general state, partially mixed$-$partially entangled state, can be
readily generated as well.
\end{abstract}


\maketitle

\textit{Introduction} $-$ Multi-particle quantum entanglement traditionally has
been associated with many fundamental issues in quantum physics, such as, the
uncertainty principle and the locality, reality and causality problems in
quantum theory \cite{eprb}.  Recently, quantum entanglement has found its
 applications in metrology, communication, and information processing
\cite{steane}. To successfully implement these new ideas, one must be able to
generate and manipulate entangled states at will. It is, however, generally
recognized that even the generation of multi-particle entanglement is not
trivial. Nonetheless, a great deal of work has been carried out with entangled
two-qubit quantum states, or Bell-states. These states are important not only
because of their simplicity, but also because of their utility in applications,
such as, quantum cryptography. In addition, entangled two-qubit states may one
day serve as building blocks for the construction of states of three or more
entangled qubits.

The first direct generation of an entangled two-qubit state involved photon
pairs produced in the process of cw-pumped type-II spontaneous parametric
down-conversion (SPDC) \cite{note11,kwiat}. Although this method is still
widely used for the generation of polarization entangled states, it has its
limitations. In particular, the photon pair emission times are completely
random. This is a drawback in applications such as quantum teleportation,
multi-photon state generation, practical quantum cryptography, etc, where
knowledge of the approximate times of emission are required.

Much of the uncertainty in emission time is eliminated when the SPDC process is
pumped by an ultrafast laser. Unfortunately, differences in the spectral and
temporal properties of the photon pair cause the polarization entanglement to
suffer with this type of pumping scheme \cite{pulsedspdctheory,pulsedspdcexp}.
It is possible to ``concentrate'' (following the definition in
Ref.~\cite{munro}) the entanglement  by passing the photons through narrow
spectral filters, effectively retaining only the more highly entangled pairs.
The entanglement concentration based on a local filtering process such as this
is not desirable, however, since most of the photons are simply wasted. A
number of schemes involving multiple crystals have been devised to circumvent
these problems \cite{kim3}, although none can match the simplicity and
stability of a single-crystal scheme.

In this paper, we report the experimental demonstration of general entanglement
concentration scheme in a two-qubit state of type-II SPDC. Our entanglement
concentration scheme does not rely on local filtering. Therefore, the degree of
entanglement is not affected by the pump bandwidth, the thickness of nonlinear
crystal, the bandwidth of spectral filters, etc. As a result, no photons are
wasted: all qubit pairs, which are initially in a mixed state, exit the
entanglement concentrator as entangled qubit pairs.

\begin{figure}[b]
\includegraphics[width=3.0in]{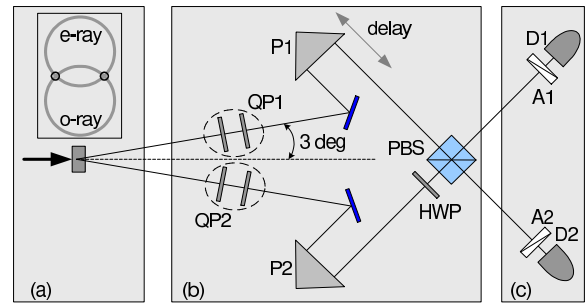}
\caption{\label{fig:setup}(a) Non-collinear type-II SPDC is used to prepare an
initial two-qubit mixed state. (b) Entanglement concentration scheme. (c)
Detectors and polarization analyzers.}
\end{figure}

\textit{Entanglement Concentration Scheme} $-$ Consider the polarization state
of the photon pair generated from a type-II BBO crystal pumped either by a cw
or by a ultrafast pump laser, see Fig.~\ref{fig:setup}(a). As in
Ref.~\cite{kwiat}, attention is restricted to the intersections of the cones
made by the e- and o-rays exiting the BBO crystal. In each of these two
directions, a photon of either polarization (horizontal or vertical) may be
found, with the orthogonal polarization found in the conjugate beam. Unlike
common misconception, the photon pairs found in these two directions are not
polarization entangled. In fact, the polarization state of the photon pair is
in a mixed state
$
\rho_{mix} = \frac{1}{2}\left(|H_1\rangle|V_2\rangle\langle V_2|\langle H_1| +
|V_1\rangle|H_2\rangle\langle H_2|\langle V_1| \right),
$
where $|H\rangle$ and $|V\rangle$ refer to the horizontal and vertical
polarization state of a single photon, respectively.

The reason the state is mixed has to do with timing information carried by the
photon pair. Because of the group velocities experienced by the different
polarizations are not the same, one polarization always precedes the other.
Thus, the two amplitudes $|H_1\rangle|V_2\rangle$ and $|V_1\rangle|H_2\rangle$
are distinguishable in principle. In cw-pumped type-II SPDC, a pair of
birefringent compensators, with the effective thickness equal to half the
down-conversion crystal, can remove this timing information, thus transforming
the mixed state to a pure polarization entangled state \cite{kwiat,cwspdc}.

But in ultrafast type-II SPDC, the pump pulse introduces additional timing
information, which cannot be eliminated with the birefringent compensators,
into the two-photon state \cite{pulsedspdctheory}. It is therefore impossible
to transform the mixed state into a pure polarization entangled state
\cite{pulsedspdcexp}.

\begin{figure}[t]
\includegraphics[width=3.0in]{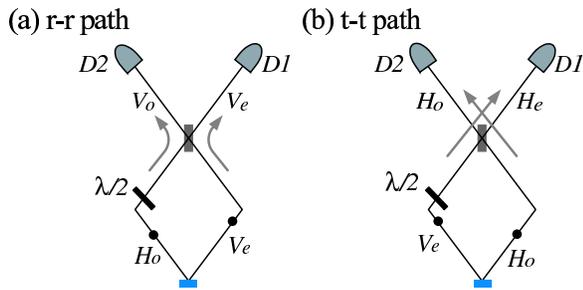}
\caption{\label{fig:feynman}Two possible quantum mechanical paths that a photon
pair may follow.}
\end{figure}

The interferometric entanglement concentration scheme shown in
Fig.~\ref{fig:setup}(b), nevertheless, allows us to transform the mixed state
to a polarization entangled state, regardless of the pump bandwidth, crystal
thickness, and spectral filters used \cite{note3}.

Let us now discuss the entanglement concentration scheme in detail, see
Fig.~\ref{fig:setup}(b). The photon pairs exit the crystal and travel equal
distances (delay $\tau=0$) to a polarization beam splitter (PBS). A $\lambda/2$
plate inserted in one arm rotates the polarization by $90^\circ$ and ensures
that the photon pairs have the same polarization when they reach the PBS. Thus,
there are two possible outcomes: both photons are reflected(r-r); or both are
transmitted (t-t). These two biphoton paths are illustrated in
Fig.~\ref{fig:feynman}. Since it is never the case that two photons exit the
same port of the PBS, no post-selection is required. Note also that the photon
that left the crystal with e-polarization (o-polarization) is always detected
by $D_1$ ($D_2$). Therefore, the intrinsic timing information present in
type-II SPDC cannot be used to distinguish between the t-t and the r-r paths.
As a result, the temporal and spectral differences between the photon pair have
no bearing on the polarization entanglement. We, therefore, have effectively
`dis-entangled' the timing information and the polarization information of the
qubit pair. Such dis-entanglement of decoherence causing and the information
carrying parts of the wavefunction of the qubit pair is the hallmark of this
entanglement concentration scheme. (Note that the photon pair arrival times are
still known within the coherence time of the pump pulse.)

The resulting amplitudes $|H_1\rangle|H_2\rangle$ and $|V_1\rangle|V_2\rangle$
are then quantum mechanically indistinguishable. Therefore, they are in quantum
superposition $ |\Phi\rangle = \frac{1}{\sqrt{2}}\left(|H_1\rangle|H_2\rangle +
e^{i\varphi} |V_1\rangle|V_2\rangle\right), $ where $\varphi$ is the phase
between the two term and it may be varied by tilting the phase plates QP1 and
QP2. With a setting of $\varphi=0$, the density matrix of the output state can
be written as $\rho_{ent} = |\Phi^{(+)}\rangle\langle\Phi^{(+)}|$.

If the delay $\tau$ is introduced in one arm relative to the other, then the
overall overlap between the two amplitudes becomes smaller. In this case, a
more general state, partially mixed$-$partially entangled state, $ \rho =
\varepsilon \rho_{ent} + (1-\varepsilon)\rho_{mix}, $ where $0 \leq \varepsilon
\leq 1$, is generated.

\textit{Universal Bell-state Synthesizer} $-$ To understand why this
entanglement concentration scheme works as a universal Bell-state synthesizer,
it is necessary to carry out fully quantum mechanical calculation of the joint
detection rate. What we would like to show here is that the quantum
interference at $\tau=0$ (balanced interferometer) does not depend on any
parameters that are related to the pump source, spectral filtering, and the
crystal properties. We present a brief summary of the calculation here and a
detailed calculation will be published elsewhere \cite{elsewhere}.

The coincidence counts $R_c$ is proportional to
$$
R_c \propto \int dt_1 dt_2|\langle0|E_1^{(+)}(t_1,\tau)
E_2^{(+)}(t_2,\tau)|\psi\rangle|^2
$$
where $|\psi\rangle$ is the state of type-II SPDC \cite{pulsedspdctheory}.
Assuming that the quartz phase plates are adjusted so that $\varphi=0$, the
electric field operators that reach the detectors in this experiment can be
written as
\begin{eqnarray}
E_1^{(+)}(t_1,\tau)&=&\int d\omega' \, \{\cos\theta_1 \,
e^{-i\omega'(t_1+\tau)}a_{Ve}(\omega')\nonumber\\&&-\sin\theta_1 \,
e^{-i\omega't_1}a_{He}(\omega')
\},\nonumber\\
E_2^{(+)}(t_2,\tau)&=&\int d\omega' \, \{\cos\theta_2 \,
e^{-i\omega't_2}a_{Vo}(\omega')\nonumber\\&&-\sin\theta_2 \,
e^{-i\omega'(t_2+\tau)}a_{Ho}(\omega') \},\nonumber
\end{eqnarray}
where, for example, $a_{Vo}(\omega')$ is the annihilation operator for a photon
of frequency $\omega'$ with vertical polarization which was originally created
as the o-ray of the crystal. $\theta_1$ and $\theta_2$ are the angles of the
polarization analyzers $A_1$ and $A_2$, respectively.

Upon carrying out the calculation, we find that the coincidence count rate has
the form
\begin{eqnarray}
R_c \propto \int dt_+ dt_-|\cos\theta_1\cos\theta_2\Pi(t_+ + \tau/2,t_- + \tau)\nonumber\\
+\sin\theta_1\sin\theta_2\Pi(t_+ + \tau/2,t_- - \tau)|^2\label{eq:coinc}.
\end{eqnarray}
The detailed expression of $\Pi(t_+,t_-)$ is given in
Ref.~\cite{kim3,elsewhere} and it contains all the important factors such as
crystal length, pump bandwidth, bandwidth of spectral filters, etc.

Let us now look at Eq.~(\ref{eq:coinc}) carefully. If the interferometer is
balanced, i.e. $\tau=0$, $\Pi(t_+,t_-)$ factors out and becomes a constant.
Therefore, $R_c\propto \cos(\theta_1-\theta_2)^2$, a clear signature of
polarization entangled state $|\Phi^{(+)}\rangle$. This means that, when
$\tau=0$, all parameters that affect the temporal shape of the $\Pi(t_+,t_-)$,
including the pump bandwidth, the crystal properties, the crystal thickness,
and the filter bandwidth, simply do not have any effect on the quantum
interference. Due to this reason, this scheme can be considered as a universal
Bell-state synthesizer.

\begin{figure}[t]
\includegraphics[width=3.0in]{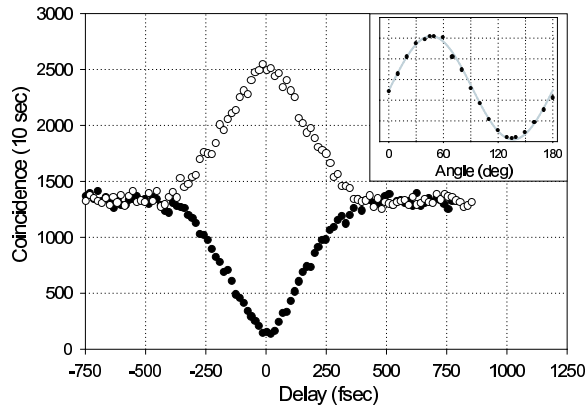}
\caption{\label{fig:cw} Experimental data for cw-pumped type-II SPDC.
$\varphi=0$ for open circle ($|\Phi^{(+)}\rangle$) and $\varphi=\pi$ for solid
circle ($|\Phi^{(-)}\rangle$). Inset shows the polarization interference for
$|\Phi^{(+)}\rangle$ state. }
\end{figure}

\textit{Experiment} $-$ As described above, the initial polarization state,
 a mixed state, was prepared by type-II non-collinear
frequency-degenerate SPDC. A 3 mm BBO crystal was pumped by a cw argon ion
laser operating at 351.1 nm, producing photons with a central wavelength of
702.2 nm. The delay $\tau$ was introduced through the motion of one of the
trombone prisms and the phase between the two alternatives (shown in
Fig.~\ref{fig:feynman}) was adjusted by slightly tilting the 600 $\mu$m quartz
plates QP1 in opposite directions. The optic axes of the quartz crystals were
oriented vertically. The outputs from the two detectors were fed to a
time-to-amplitude converter (TAC) and the TAC output was analyzed by a
multi-channel analyzer with a coincidence window set to 3 nsec. No spectral
filters were used in this experiment.

First, we have done typical space-time interference experiments by varying the
delay $\tau$ with both polarization analyzers set at $45^\circ$. The phase term
was adjusted by tilting the quartz plates QP1 and $\varphi=0$ and $\varphi=\pi$
were selected to prepare Bell-states $|\Phi^{(+)}\rangle$ and
$|\Phi^{(-)}\rangle$, respectively. Fig.~\ref{fig:cw} shows the experimental
data. When the delay $\tau$ is zero, i.e., no path length difference between
the two arm, complete destructive or constructive interference is observed.
Note that this is different from that of typical type-II case in which $\tau$
equal to half the crystal thickness should be inserted to observe complete
quantum interference \cite{kwiat,cwspdc}. The typical triangular two-photon
wavepacket is clearly demonstrated and the base width of the triangular
wavepacket agrees well with the theoretically expected value of 742 fsec
\cite{cwspdc,elsewhere}.

The inset of Fig.~\ref{fig:cw} shows the polarization interference for
$|\Phi^{(+)}\rangle$ state. $A_1$ was fixed at $45^\circ$ and $A_2$ was
rotated. Expected $\cos(\theta_1-\theta_2)^2$ correlation is clearly
demonstrated.

We have repeated the above space-time and polarization measurements with
several different spectral filters and found almost no change in the quality of
quantum interference. The stability of the interferometer is also checked by
repeating the polarization interference measurement several times at different
times and we have found almost no change in the visibility of the polarization
interference. This is basically due to the fact that a small change around
$\tau=0$ causes very little effects on the quantum interference as shown in
Fig.~\ref{fig:cw}.

\begin{figure}[t]
\includegraphics[width=3.0in]{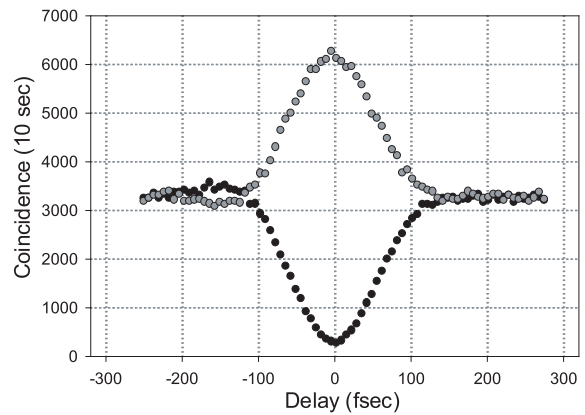}
\caption{\label{fig:pulse} Experimental data for ultrafast type-II SPDC.
Analyzer angles were $45^\circ$/$45^\circ$ for the peak and
$45^\circ$/$-45^\circ$ for the dip.}
\end{figure}

We have also done the experiment in which an ultrafast laser is used as the
pump. A second harmonic (390 nm) of a commercial mode-locked Ti:Sa laser
operating at 780 nm was used as a pump. The bandwidth of the pump pulse was
about 2 nm. The thickness of the type-II BBO crystal was 3 mm and the SPDC
radiation was centered at 780 nm. Interference filters with 20 nm FWHM were
inserted in front of each detectors to reduce noise counts. The effective
coincidence window for this experiment was 3 nsec. The quartz phase plates QP1
and QP2 were not available for this experiment so the phase $\varphi$ could
only be roughly adjusted by the tilt angle of the crystal itself. Since the
phase $\varphi$ could not be freely adjusted, we fixed $\varphi$ at some
arbitrary angle and scanned the delay $\tau$ for analyzer angles
$45^\circ$/$45^\circ$ for the peak and $45^\circ$/$-45^\circ$ for the dip.
Fig.~\ref{fig:pulse} shows a typical data set for this measurement. The
observed visibility is greater than 90\% and could have been improved further
if phase plates had been available to fine tune $\varphi$. Nevertheless, the
observed visibility is by far the highest in ultrafast type-II SPDC generated
from a thick crystal (For comparison, see Ref.~\cite{pulsedspdcexp} for usual
ultrafast type-II SPDC data) and in principle it does not depend on crystal,
pump, and filter parameters as shown in Eq.~(\ref{eq:coinc}). The FWHM of the
interference is roughly estimated to be 140 $\sim$ 150  fsec which is very
close to theoretically expected value of 160 fsec (assuming exactly 2 nm FWHM
pump bandwidth). \cite{elsewhere}.

\textit{Discussion} $-$ It is helpful to introduce the terms ``Entanglement''
or ``Entanglement of Formation'' ($E$) and ``Entropy'' or ``Entropy of
Entanglement'' ($S$) formally, as done in Ref.~\cite{munro}, to visualize the
entanglement concentration process in this work.

Under these definitions, our initial mixed two-qubit state has $S=0.5$ and
$E=0$ \cite{init}. The output state of the balanced entanglement concentrator
has $E=1$ and $S=0$. Most importantly, the entanglement concentration has been
made without throwing away any sub-ensemble of the initial two-qubit system.

A more general state, partially mixed$-$partially entangled, can be readily
prepared by introducing the delay $\tau$ in one arm. If $\tau \neq 0$ and
$\tau<|\delta t|$, where $\delta t$ is the two-photon coherence length, we
obtain a state which is not completely entangled and not completely mixed. Such
states are called Werner states \cite{werner} and lie between two points
($S$,$E$)$=$(0.5,0) and ($S$,$E$)$=$(0,1). Only recently, researchers have
started to study these states experimentally in the cw domain \cite{kwiat2}.
Such states are important in studying controllable decoherence in multi-qubit
systems (for now, however, it is limited to two-qubit systems). Since our
scheme offers readily controllable decoherence in the pulsed domain, we believe
that it will be a useful tool to generate a multi-qubit entangled state and to
study its decoherence in a controlled environment.

In conclusion, we have reported the experimental realization of entanglement
concentration without local filtering and demonstrated a universal Bell-state
synthesizer. Although typical type-II SPDC is used as a source, our Bell-state
synthesizer is not affected by the pump source, the crystal properties, and the
use of spectral filters. In addition, a more general state, partially
mixed-partially entangled state can be readily generated. We believe that this
new Bell-state synthesizer will be useful in experimental studies of quantum
information science and as a building block of multi-particle entanglement.
Finally, the entanglement concentration scheme based on dis-entanglement of
decoherence causing and information carrying parts of the wavefunction may be
applied to other quantum systems in which strong decoherence in one variable
destroys entanglement in the other variable.

We would like to thank M.H. Rubin for helpful discussions. This research was
supported by the U.S. Department of Energy, Office of Basic Energy Sciences,
NSA, NSF, and ONR. The Oak Ridge National Laboratory is managed for the U.S.
DOE by UT-Battelle, LLC, under contract No.~DE-AC05-00OR22725.


\begin{thebibliography}{}
\bibitem{note11} Here we only consider experiments in which amplitude
post-selection does not have to be assumed.
\bibitem{note3} Our entanglement concentration scheme does not remove the intrinsic
timing information present in type-II SPDC: it is simply made irrelevant to the
quantum interference effect.
\bibitem{init} $S=-\sum_{i=1}^4 \lambda_i \log_4 \lambda_i$, where $\lambda_i$
are the eigenvalues of the density matrix $\rho$. $S=0.5$ because the initial
state is a mixture of $|H_1\rangle|V_2\rangle$ and $|V_1\rangle|H_2\rangle$ of
all four possible states.
\end{thebibliography}
\end{document}